\documentclass[twocolumn,aps,prc,floatfix]{revtex4}
\usepackage{graphicx}
\usepackage{dcolumn}
\usepackage{bm}
\usepackage[version=4]{mhchem}
\usepackage{color}
\def\esym{$E_{\rm sym}(\rho)$~}

\def\es0{$E_{\rm sym}(\rho_0)$}

\def\us0{$U_{\rm sym}(\rho_0,k_F)$~}

\def\l0{$L(\rho_0)$~}

\renewcommand{\d}{\mathrm{d}}
\begin{document}
\title{Probing high-density nuclear symmetry energy with $\Xi^{-}/\Xi^{0}$ ratio in heavy-ion collisions at $\sqrt{s_{NN}} \sim 3$ GeV }
\author{Gao-Chan Yong$^{1,2}$\footnote{yonggaochan@impcas.ac.cn}, Bao-An Li$^3$\footnote{Bao-An.Li@Tamuc.edu}, Zhi-Gang Xiao$^4$\footnote{xiaozg@mail.tsinghua.edu.cn} and Zi-Wei Lin$^5$\footnote{LINZ@ecu.edu}}

\affiliation{$^1$Institute of Modern Physics, Chinese Academy of Sciences, Lanzhou 730000, China}
\affiliation{$^2$School of Nuclear Science and Technology, University of Chinese Academy of Sciences, Beijing, 100049, China}
\affiliation{$^3$Department of Physics and Astronomy, Texas A$\&$M
University-Commerce, Commerce, TX 75429-3011, USA}
\affiliation{$^4$Department of Physics, Tsinghua University, Beijing 100084, China}
\affiliation{$^5$Department of Physics, East Carolina University, Greenville, North Carolina 27858, USA}

\setcounter{MaxMatrixCols}{10}
\begin{abstract}
Recent beam energy scan (BES) experiments at RHIC by the STAR Collaboration (PLB {\bf 827},137003 (2022) and PRL {\bf 128}, 202303 (2022)) found that hadronic interactions dominate the collective flow and the proton cumulant ratios
are driven by baryon number conservation in a region of high baryon density in $\sqrt{s_{NN}}$ = 3 GeV Au+Au reactions, indicating the dense medium formed in such collisions is likely hadronic matter. Within an updated ART (A Relativistic Transport) model with momentum dependent isoscalar and isovector single-nucleon mean-field potentials corresponding to different symmetry energies at suprasaturation densities, the $n/p$, $\pi^{-}/\pi^{+}$, $K_{s}^{0}/K^{+}$, $\Sigma^{-}/\Sigma^{+}$ and $\Xi^{-}/\Xi^{0}$ ratios are studied for central Au+Au collisions at $\sqrt{s_{NN}}$ = 3 GeV where the maximum central density reaches about $(3.6\sim 4.0)\rho_0$. The doubly strange $\Xi^{-}/\Xi^{0}$ ratio is found to have the strongest sensitivity to the variation of high-density nuclear symmetry energy. Thus, the $\Xi^{-}/\Xi^{0}$ ratio in relativistic heavy-ion reactions at $\sqrt{s_{NN}} \sim 3$ GeV may help probe sensitively the poorly known symmetry energy of dense neutron-rich matter critically important for understanding various properties of neutron stars.
\end{abstract}
\maketitle

%\linenumbers
%\twocolumn
\section{Introduction}
Nuclear symmetry energy \esym at density $\rho$ measures the energy cost to make nuclear matter more neutron rich. It is critically important for understanding the evolution of QCD phase diagram with isospin asymmetry in general and properties of neutron stars in particular \cite{esymreview}. Indeed, much progress has been achieved in constraining the \esym around and below the saturation density $\rho_0$ of nuclear matter using combined information from nuclear theories, terrestrial nuclear experiments and observations of neutron stars using various messengers over the last two decades, see, e.g., refs.~\cite{Li98,Ste05,Baran,BALI,Tsang12,Trau12,Baldo16b,Li17,LCCX18,Yong20,epja19,Burg21} for reviews. The \esym is normally Taylor expanded or simply parameterized as
\begin{eqnarray}
E_{\rm{sym}}(\rho)&=&E_{\rm{sym}}(\rho_0)+L(\frac{\rho-\rho_0}{3\rho_0})\nonumber\\
&+&\frac{K_{\rm{sym}}}{2}(\frac{\rho-\rho_0}{3\rho_0})^2+\frac{J_{\rm{sym}}}{6}(\frac{\rho-\rho_0}{3\rho_0})^3
\end{eqnarray}
where the magnitude $E_{\rm{sym}}(\rho_0)$, slope parameter $L=[3\rho\d E_{\rm{sym}}/\d\rho]_{\rho_0}$, curvature parameter $K_{\rm{sym}}=[9\rho^2\d^2 E_{\rm{sym}}/\d\rho^2]_{\rho_0}$, and skewness
parameter $J_{\rm{sym}}=[27\rho^3\d^3 E_{\rm{sym}}/\d\rho^3]_{\rho_0}$ at the saturation density $\rho_0$ of nuclear matter are used to characterize the density dependence of nuclear symmetry energy.
Quantitatively, surveys of about 100 analyses of various data in the literature found that most of them gave $E_{\rm sym}(\rho_0)$ and $L$  scatter around $E_{\rm sym}(\rho_0)=31.6\pm 2.7$ MeV \cite{Li2013,as5} and
$L=57.7\pm 19$ MeV \cite{Li2013,as5,LiBA21} with some exceptions, respectively.  However, determining the high-density behavior of \esym has been found extremely challenging.
On one hand, it is very encouraging to see that the magnitude of \esym at $2\rho_0$ from analyses of heavy-ion reaction experiments \cite{Rus11,Rus16} and studies \cite{Zhang18,France1,Zhang19,as6,Nakazato19,Xie19,Tsang20,Xie20,Tan20,XZhang20,Zhang21,Mondal22,Jerome22,Yue21} of multimessengers from recent neutron star observations \cite{LIGO18,Miller21,Riley21} as well as calculations based on the state-of-the-art nuclear many-body theories employing chiral effective interactions \cite{Ohio20,Diego} start to merge around $E_{\rm sym}(2\rho_0)\approx 51\pm 13$ MeV \cite{LiBA21}. On the other hand, theoretical predictions for \esym at higher densities diverge \cite{epja19} and the results \cite{Li17,Zhang21,Mondal22} extracted from both heavy-ion reactions and the latest observation of PSR J0740+6620 with mass $2.08\pm 0.07~M_{\odot}$ by NICER spans from being super-stiff (the \esym increases faster than $\rho$) to super-soft (quickly decreases to zero).
Quantitatively, the $K_{\rm{sym}}$ characterizing the \esym around $2\rho_0$ from 16 new analyses of neutron star observables was determined to be about $K_{\rm{sym}}\approx -107\pm 88$ MeV while the $J_{\rm{sym}}$ characterizing the \esym above about $(2.5\sim 3.0)\rho_0$ is largely unconstrained \cite{LiBA21}. In fact, many interesting physics questions regarding the high-density \esym as well as the possible hadron-quark phase transition, fundamental symmetries and kaon condensation in the cores of neutron stars are strongly intertwined \cite{Kut94,Kut98,Kubis1,Rho}. Thus, the high-density behavior of \esym has long been regarded as one of the most uncertain parts of the dense neutron-rich matter Equation of State (EOS) \cite{Kut94,Li2002}. Its determination is expected to have significant impact on understanding the QCD phase diagram of dense neutron-rich matter.

Indeed, there are strong interests in probing the high-density \esym as an integral part of the science missions of several relativistic heavy-ion reaction facilities under construction, see, e.g, Refs. \cite{Xiao14,Wolfgang19,GSI,Senger} for preliminary plans to carry out such studies at HIAF/China, FAIR/Germany and NICA/Russia. One of the critical prerequisites for realizing this scientific goal is  to find potential observables that are sensitive to the variation of high-density \esym within its currently known uncertainty range. The isospin asymmetry of nucleonic component of dense matter formed in heavy-ion reactions varies with beam energy and once the quark deconfinment happens the \esym loses its physical meaning. It is thus particularly useful to probe the high-density \esym in heavy-ion reactions that create hadronic matter as dense as possible but not high enough to form the quark-gluon plasma (QGP) yet. Of course, one wishes to make the dense matter as neutron rich as possible using probably high-energy radioactive beams. Very interestingly, the STAR Collaboration found recently that the quark-scaling signature in collective flow disappears and all collective flow data can be well described by hadronic transport models using nuclear mean-field potentials in $\sqrt{s_{NN}}$ = 3 GeV Au+Au reactions. Moreover, the measurements of proton high-order cumulants imply that the QCD critical region, if created in heavy-ion collisions, could only exist at energies higher than 3 GeV. These findings imply that the dense medium formed in $\sqrt{s_{NN}}$ = 3 GeV Au+Au collisions is likely hadronic matter \cite{casplb22,proprl22}.

In this work, we explore in the aforementioned reactions studied by STAR potential observables for studying the high-density \esym using the ART model \cite{ART1,art2001}. The latter has been extensively used and continuously improved in several aspects \cite{ko2002npa,ko2004npa,ko2004plb,chen2004plb,li2012prc,deu2009} over the years as the hadronic afterburner in the publicly available AMPT (A Multiphase Transport) package \cite{AMPT2005} for simulating relativistic heavy-ion collisions from RHIC-BES to LHC energies, see Ref. \cite{Lin21-review} for a recent review. To investigate effects of high-density \esym on relativistic heavy-ion reactions, we incorporated in the ART model the momentum dependent isoscalar and isovector single-nucleon mean-field potentials \cite{das2003} that have been previously used extensively by some of us in studying many topics of isospin physics in nuclear structures, low-intermediate energy heavy-ion reactions and properties of neutron stars \cite{BALI}. We found that the doubly strange $\Xi^{-}/\Xi^{0}$ ratio is the most sensitive observable to the variation of  high-density \esym compared to the $n/p$, $\pi^{-}/\pi^{+}$, $K_{s}^{0}/K^{+}$, and $\Sigma^{-}/\Sigma^{+}$ ratios in $\sqrt{s_{NN}}$ = 3 GeV Au+Au reactions.

\section{An updated ART model within AMPT}
The ART afterburner in the latest AMPT package simulates the dynamical evolution of the phase space distribution functions of
nucleons, $\Delta(1232)$, $N^*(1440)$, $N^*(1535)$, $\Lambda$, $\Sigma$, $\Xi$, $\Omega$, $\pi$, $\rho$, $\omega$, $\eta$, $K$, $K^*$, $\phi$ and deuteron. Many scatterings among these particles and the corresponding inverse reactions are included, see Refs. \cite{AMPT2005,deu2009} for details. In the present work, the Woods-Saxon nucleon density distribution and local Thomas-Fermi approximation are used to initialize the position and momentum of each nucleon in the projectile and target \cite{impbuu}. As in the original ART code, the test-particle method is used to evaluate the density matrices and mean-field potentials for nucleons, baryonic resonances, $K$, $\Lambda$, $\Sigma$, $\Xi$ and $\Omega$ as well as their antiparticles.
For studying isospin physics especially the high-density \esym with relativistic heavy-ion collisions, we adopt the following isospin and momentum dependent nucleon potential \cite{das2003}
\begin{eqnarray}
&&U(\rho,\delta,\vec{p},\tau)=A_u(x)\frac{\rho_{\tau'}}{\rho_0}+A_l(x)\frac{\rho_{\tau}}{\rho_0}\nonumber\\
& &+B(\frac{\rho}{\rho_0})^{\sigma}(1-x\delta^2)-8x\tau\frac{B}{\sigma+1}\frac{\rho^{\sigma-1}}{\rho_0^\sigma}\delta\rho_{\tau'}\nonumber\\
& &+\frac{2C_{\tau,\tau}}{\rho_0}\int
d^3\,\vec{p^{'}}\frac{f_\tau(\vec{r},\vec{p^{'}})}{1+(\vec{p}-\vec{p^{'}})^2/\Lambda^2}\nonumber\\
& &+\frac{2C_{\tau,\tau'}}{\rho_0}\int
d^3\,\vec{p^{'}}\frac{f_{\tau'}(\vec{r},\vec{p^{'}})}{1+(\vec{p}-\vec{p^{'}})^2/\Lambda^2},
\label{buupotential}
\end{eqnarray}
where $\rho_0$ denotes the saturation density, $\tau, \tau' = 1/2(-1/2)$ for neutron (proton).
The parameter $x$ was introduced to mimic different forms of \esym predicted by various nuclear
many-body theories without changing any property of the symmetric nuclear matter and the magnitude of \esym at $\rho_0$.
It is a phenomenological potential based on Gogny-Hartree-Fock (GHF) predictions with parameters constrained by empirical properties of nuclear matter at $\rho_0$ as well as isoscalar and isovector single-nucleon optical potentials at $\rho_0$, see, e.g., Refs. \cite{BALI,Chen14} for reviews. With $x=1$, the single-particle potential and the corresponding EOS reproduce the default GHF prediction \cite{das2003}. The above single-particle potential has been used previously in the isospin-dependent Boltzmann-Uehling-Uhlenbeck (IBUU) transport model for studying isospin physics with low-intermediate energy heavy-ion reactions \cite{LiBA04,LiChen05}. IBUU analyses of isospin diffusion experiments in Sn+Sn reactions at $E_{\rm beam}$/A= 50 MeV indicates that the parameter x is between x=0 and x=-1 at densities below about $1.2\rho_0$ \cite{Chen-prl} while analyses of the $\pi^-/\pi^+$ ratio from Au+Au reactions at $E_{\rm beam}$/A =$400\sim 1000$ MeV prefer $x=1$ leading to a super-soft \esym at densities around $(1.5-2.5)\rho_0$ \cite{Xiao-prl}. On the other hand, some other observables in similar reactions have found indications that the \esym around $(1.5-2.5)\rho_0$ might be stiff and closer to that with $x=-1$ \cite{Rus11,Rus16}. In this exploratory study, we thus use $x=1$ and $x=-1$ as two limiting cases. The corresponding two \esym functions are compared in the upper inset of Fig.1. Specific values of other parameters used in this work in the single-particle potentials of Eq.(\ref{buupotential}) can be found in Ref.~\cite{impbuu}.

The form of kaon potential was taken from Ref.~\cite{ligq97} while no mean-field potential is used for pions. For strange baryons $\Lambda$, $\Sigma$, $\Xi$, we adopt the quark counting rule asserting that these strange baryons interact with other baryons only through their non-strange (2/3, 2/3, 1/3) constituents \cite{chung2001,mos74}. 
Moreover, we use the known decay branching ratios of different isospin multiplets of these strange baryons to determine the relationships of their mean-field potentials with those for neutrons and protons. Putting these assumptions together, we have
\begin{eqnarray}
U_{\Lambda} &=& 2/3(1/3U_{n}+2/3U_{p}), \nonumber\\
U_{\Sigma^{-}} &=& 2/3U_{n}, \nonumber\\
U_{\Sigma^{0}} &=& 2/3(1/3U_{n}+2/3U_{p}), \nonumber\\
U_{\Sigma^{+}} &=& 2/3(1/2U_{n}+1/2U_{p}), \nonumber\\
U_{\Xi^{-}} &=& 1/3(1/3U_{n}+2/3U_{p}), \nonumber\\
U_{\Xi^{0}} &=& 1/3(1/3U_{n}+2/3U_{p}).
\end{eqnarray}
To describe the doubly strange $\Xi$ production at low energies, besides the strangeness exchange reactions $\bar{K}$+$Y$ $\leftrightarrow$ $\pi$+$\Xi$ (Y = $\Lambda$ or $\Sigma$) that are already in the AMPT package, the isospin-averaged cross sections are used for simulating the $Y$+$Y$ $\leftrightarrow$ $N$+$\Xi$ reactions and $\Xi$ productions via the $Y$+$N$ $\rightarrow$ $N$+$\Xi$+$K$ processes \cite{li2012prc,urqmd2014,urqmd2016,buu2018}.

\section{Probing high-density symmetry energy with hadron isospin multiplet ratios}
It is known that the strength of isovector interactions are relatively weak compared to the isoscalar ones and the isospin asymmetry reachable in heavy-ion reactions is small. Moreover, theoretically there are still many unknowns about the isovector nuclear interaction especially at suprasaturation densities. Thus, the study of isospin physics in heavy-ion reactions and the search of their experimental signatures are currently suffering from some essentially unavoidable model dependences. Nevertheless, it is interesting to note that some transport model developers and users have been carrying out systematically the Transport Model Evaluation Project over the last few years to better understand and possibly reduce the model dependencies, see, e.g, Ref. \cite{TMEP} for the most recent review of these efforts. Certainly, there is a strong interest in the community to find strong and clean experimental observablers sensitive to the variation of \esym at suprasaturation densities.

Since strange mesons or baryons are rarely absorbed by the surrounding medium, they have long been recognized as particular useful probes of the dense matter EOS \cite{raf82,jor1985,fuchs2001,kaon2006,ditoro,fuchs2006,feng11,liq05,qm2018,adam20,casyong2021}.
In particular, kaons have been thoroughly studied in the literature \cite{fuchs2001,kaon2006,ditoro,fuchs2006,jor1985,feng11}. The singly strange $\Lambda$ and $\Sigma$ hyperons have been studied by the E895 and FOPI Collaborations \cite{chung2001}, and their connections to the nuclear EOS were explored theoretically \cite{liq05,fengnpa2013,fengcpl2021}. The doubly strange $\Xi$ production in heavy-ion collisions has also been continuously studied over the last twenty years \cite{exp2003,exp2009,exp2015,ko2002npa,ko2004npa,ko2004plb,chen2004plb,li2012prc,urqmd2014,urqmd2016,buu2018}.
The $\Xi$ hyperon comes mainly from collisions of two singly strange particles. Its fraction in the central cell of the participant region is more than twice that of $K^{+}$ or $\Lambda+\Sigma^{0}$ \cite{casyong2021}. It may thus have a significant advantage over singly strange hadrons in probing the EOS of dense matter \cite{casyong2021}. However, its dependence on the EOS is less known and the elementary cross sections of its reaction channels suffer from larger uncertainties. To probe the \esym at suprasaturation densities, we shall focus on the $\Xi^{-}/\Xi^{0}$ ratio in comparison with the isospin multiplet ratios of other particles, namely, the $n/p$, $\pi^{-}/\pi^{+}$, $K_{s}^{0}/K^{+}$, and $\Sigma^{-}/\Sigma^{+}$ ratios. These ratios will naturally reduce effects of uncertainties due to both the symmetric nuclear matter EOS and the elementary reaction cross sections. We notice that some of these ratios have been used previously in probing the high-density \esym in heavy-ion collisions at lower beam energies, see, e.g. Ref. \cite{BALI} for a review.

To understand qualitatively why the ratios of isospin multiplets mentioned above are useful for probing the $E_{\rm sym}(\rho)$, it is useful to first recall a few well-known relationships of isospin asymmetric nuclear matter and expectations for some of the particle ratios based on statistical models. In cold isospin asymmetric matter of isospin asymmetry $\delta\equiv (\rho_n-\rho_p)/(\rho_n+\rho_p)$, the energy density $\epsilon(\rho,\delta)\approx \rho\cdot[E_{\rm SNM}(\rho)+E_{\rm sym}(\rho)\delta^2]$ where $E_{\rm SNM}(\rho)$ is the average energy per nucleon in symmetric nuclear matter (SNM). The chemical potentials of neutrons/protons (n/p) are determined by $\mu_{n/p}=\partial\epsilon/\partial\rho_{n/p}$. The resulting difference in neutron-proton chemical potentials is $\mu_n-\mu_p=4E_{\rm sym}(\rho)\delta$, relating isospin multiplet ratios in statistical models with the symmetry energy. At a finite temperature T for finite systems, besides the above and a Coulomb potential there are thermal terms related to $\rho_n-\rho_p$ and $\rho_n/\rho_p$ in the expression of $\mu_n-\mu_p$ \cite{thermal}. These results have two important implications for the present work. Firstly, considering two connected regions at average densities $\rho_1$ and  $\rho_2$ formed during heavy-ion reactions, chemical equilibrium conditions (same chemical potentials in the two regions for both neutrons and protons) lead to $E_{\rm sym}(\rho_1)\delta(\rho_1)=E_{\rm sym}(\rho_2)\delta(\rho_2)$. It is the physics origin of the so-called isospin fractionation, namely, in a region where the \esym is higher, the isospin asymmetry $\delta$ there will be lower. For example, in the liquid-gas phase transition region since the \esym continuously increases with increasing density, the gas phase is expected to be more neutron-rich than the liquid phase \cite{muller,liko,baran,shi,HXu}. At suprasaturation densities, the \esym could either increase (stiff) or decrease (soft) with density, then the corresponding $\delta$ (n/p ratio) would be lower or higher.

Secondly, it is well known that the primordial $\pi^{-}/\pi^{+}$ ratio from the first-chance production channels through baryon resonances is proportional to $(N/Z)^2$ where the $N$ and $Z$ are the neutron and proton numbers of the reaction system \cite{Stock}. While in statistical models \cite{Bertsch,Aldo}, the $\pi^{-}/\pi^{+}$ ratio is proportional to ${\rm exp}\left[2(\mu_n-\mu_p)/T\right]$ which is approximately $(n/p)^2$ at freeze-out of the reaction. Thus, the $\pi^{-}/\pi^{+}$ ratio is expected to be sensitive to the density dependence of \esym \cite{Li2002}. This dependence on \esym will be carried over to the subsequent reactions involving pions. In particular, the singly strange particles are produced mostly through collisions involving energetic pions and their subsequent scatterings will produce the doubly strange particles. The ratios of the strange isospin multiplets are thus also expected to carry useful information about the $E_{\rm sym}(\rho)$. As these particles can escape with little final state interactions compared to pions, they may carry more reliable information about the \esym at suprasaturation densities. More specifically, it was shown in a thermal model for particle production in relativistic heavy-ion collisions \cite{li2012prc} that $\Xi^{-}/\Xi^{0}=\Sigma^{-}/\Sigma^{0}=\Sigma^{0}/\Sigma^{+}=N/Z={\rm exp}[(-\mu_{c})/T]$ where the $\mu_c$ is the charge chemical potential. As discussed earlier, the $\mu_c$ contains the $\mu_n-\mu_p=4E_{\rm sym}(\rho)\delta$, a Coulomb potential and thermal terms. Thus, potentially they can all carry some useful information about the high-density symmetry energy. Of course, the reality might be significantly different from the above expectations based on the physics intuitions and idealized models. Moreover, the expectations based on thermal models do not provide a rank of the \esym sensitivities of these observables. Therefore, simulations based on transport models are invaluable.

\section{ART model predictions}
\subsection{Isospin asymmetry of hadronic matter in relativistic heavy-ion collisions}
In relativistic heavy-ion collisions, the nucleonic isospin asymmetry in the projectile and target is quickly converted to isospin asymmetries of newly produced particles. As a measure of isospin asymmetry of baryonic matter we count neutron-like and proton-like particles using the branching ratios of the decay channels of baryon resonances and hyperons according to
\begin{eqnarray}
&&n_{\rm like}=n+\Delta^{-}+\Sigma^{-}+1/2\Sigma^{+} \nonumber\\
        &&+1/3(\Delta^{+}+N_{1440,1535}^{0}+\Lambda+\Sigma^{0}+\Xi^{-}+\Xi^{0}) \nonumber\\
        &&+2/3(\Delta^{0}+N_{1440,1535}^{+}), \nonumber
\end{eqnarray}
\begin{eqnarray}
&&p_{\rm like}= p+\Delta^{++}+1/2\Sigma^{+} \nonumber\\
        &&+2/3(\Delta^{+}+N_{1440,1535}^{0}+\Lambda+\Sigma^{0}+\Xi^{-}+\Xi^{0}) \nonumber\\
        &&+1/3(\Delta^{0}+N_{1440,1535}^{+}). \nonumber
\end{eqnarray}

\begin{figure}[thb]
\centering
  \resizebox{0.45\textwidth}{!}{
  \includegraphics{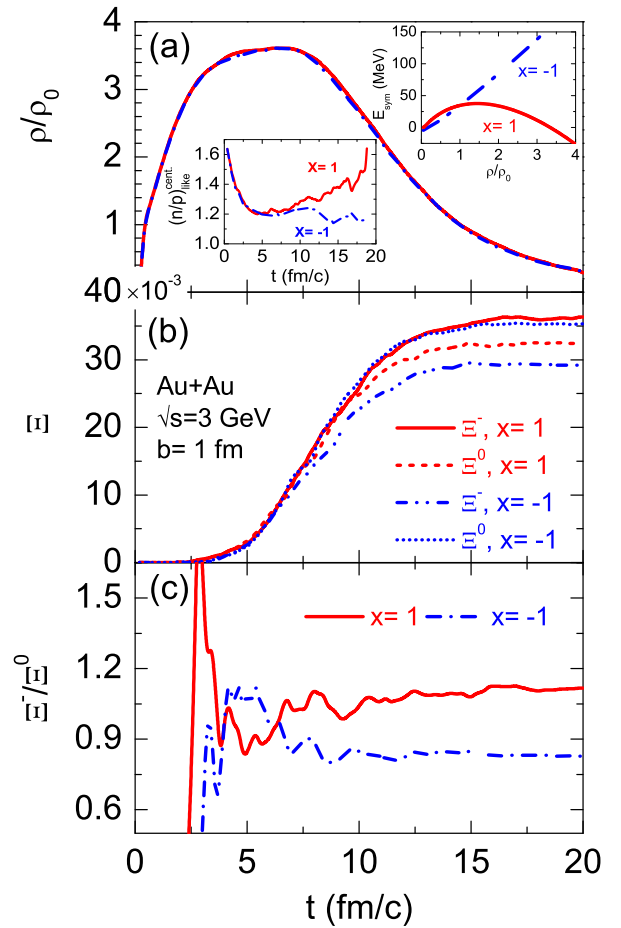}
  }
\caption{Evolutions of central compression densities (a), yields of $\Xi^{-}$ and $\Xi^{0}$ (b) and the ratios of $\Xi^{-}/\Xi^{0}$ (c) in central Au+Au reactions at $\sqrt{s_{NN}}$ = 3 GeV with soft and stiff symmetry energies. The upper-inset in panel (a) shows the density dependences of the symmetry energy with parameters $x = 1$ (soft) and $x = -1$ (stiff) in Eq.~(\ref{buupotential}). The middle-inset  in panel (a) shows the evolution of the isospin asymmetry $(n/p)_{\rm like}$ in a single central cell of 1 cubic Fermi. } \label{evo}
\end{figure}
In Fig.~\ref{evo}, the ratio $(n/p)_{\rm like}$ (the middle-inset of the upper panel) in a central cell of 1 fm$^{3}$ and the central density are shown as functions of time in the central Au+Au collisions at $\sqrt{s_{NN}}$ = 3 GeV with the two \esym functions shown in the upper-inset. It is seen that the $(n/p)_{\rm like}$ decreases quickly in the earlier phase as new particles, such as pions, carry away a large fraction of the initial isospin asymmetry of the reaction system. During this period, the $(n/p)_{\rm like}$ ratio is largely independent of the \esym used. After about 5 fm/c, appreciable \esym effects start showing up.
The $(n/p)_{\rm like}$ with the soft \esym (with x=1) is obviously higher than that with the stiff symmetry energy.
In the later stage, the central cell becomes a low density region where the isospin asymmetry becomes higher as one expects from the isospin fractionation mechanism. The maximum compression reached is about $3.6\rho_{0}$, lower than that from using a momentum-independent single particle potential with approximately the same incompressibility as the momentum-dependent part of the potential is repulsive \cite{art2001}. We notice that the symmetry energy has negligible effects on the compression as the isospin asymmetric pressure is much smaller than the SNM pressure at the high densities reached. 
There is at present a huge debate in the compact star community on the critical baryonic density above which it is realistic to suppose that quark deconfinement might occur. Vey interestingly, our results shown in the upper panel of Fig.~\ref{evo} enable us to translate the recent findings by the STAR collaboration~\cite{casplb22,proprl22} into an estimate of the lower boundary for this critical density, i.e, about $3.6\rho_{0}$.

\begin{figure}[thb]
\centering
\includegraphics[width=0.45\textwidth]{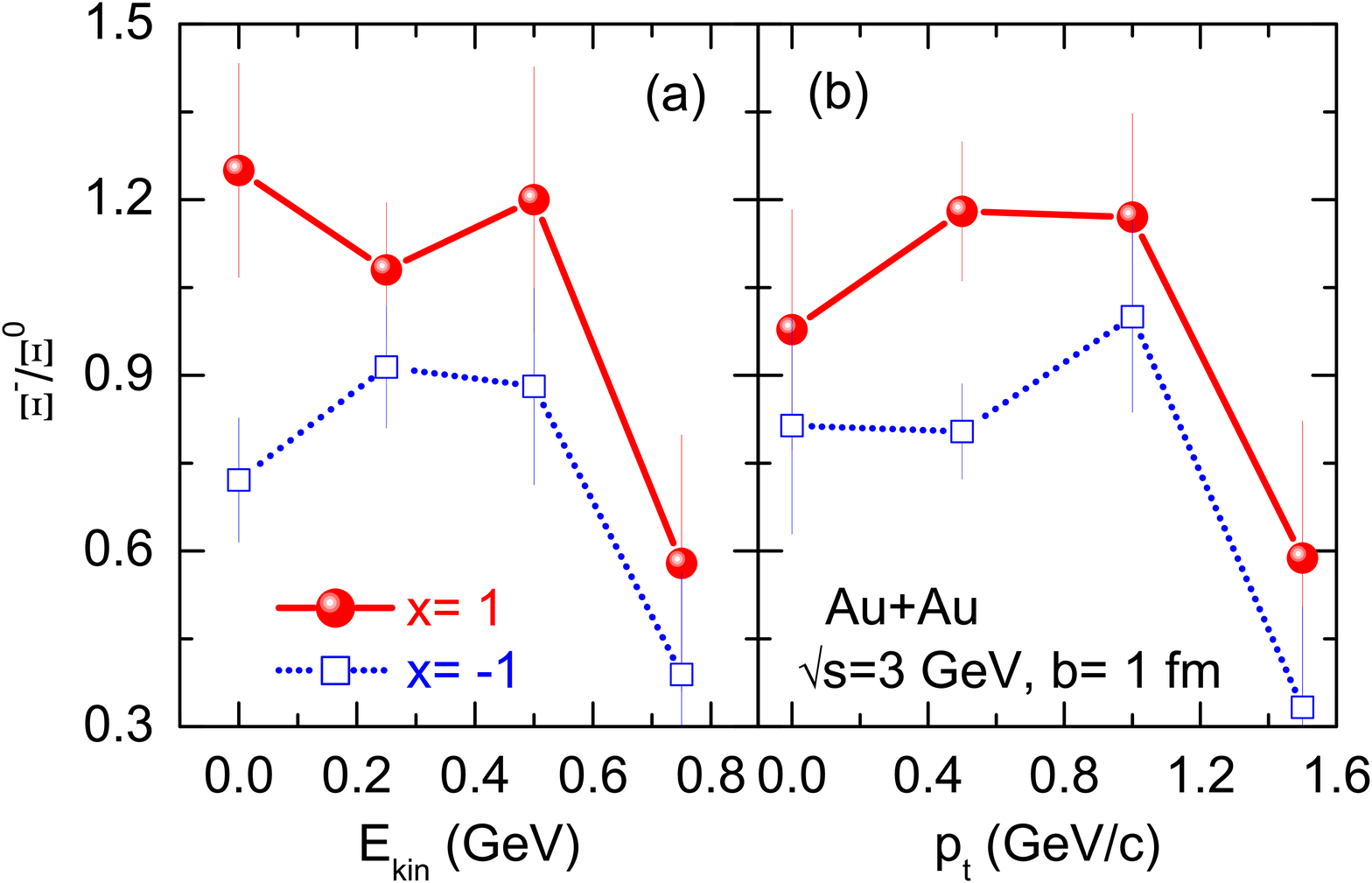}
\caption{Kinetic energy (a) and transverse momentum (b) distributions of the doubly strange baryon $\Xi^{-}/\Xi^{0}$ ratio in the central Au+Au reactions at $\sqrt{s_{NN}}$ = 3 GeV with the stiff and soft symmetry energies, respectively. The curves are used to guide the eye and the error bars are statistical in nature.} \label{rcas}
\end{figure}
\begin{figure*}
  \centering
   \resizebox{1.0\textwidth}{!}{
  \includegraphics{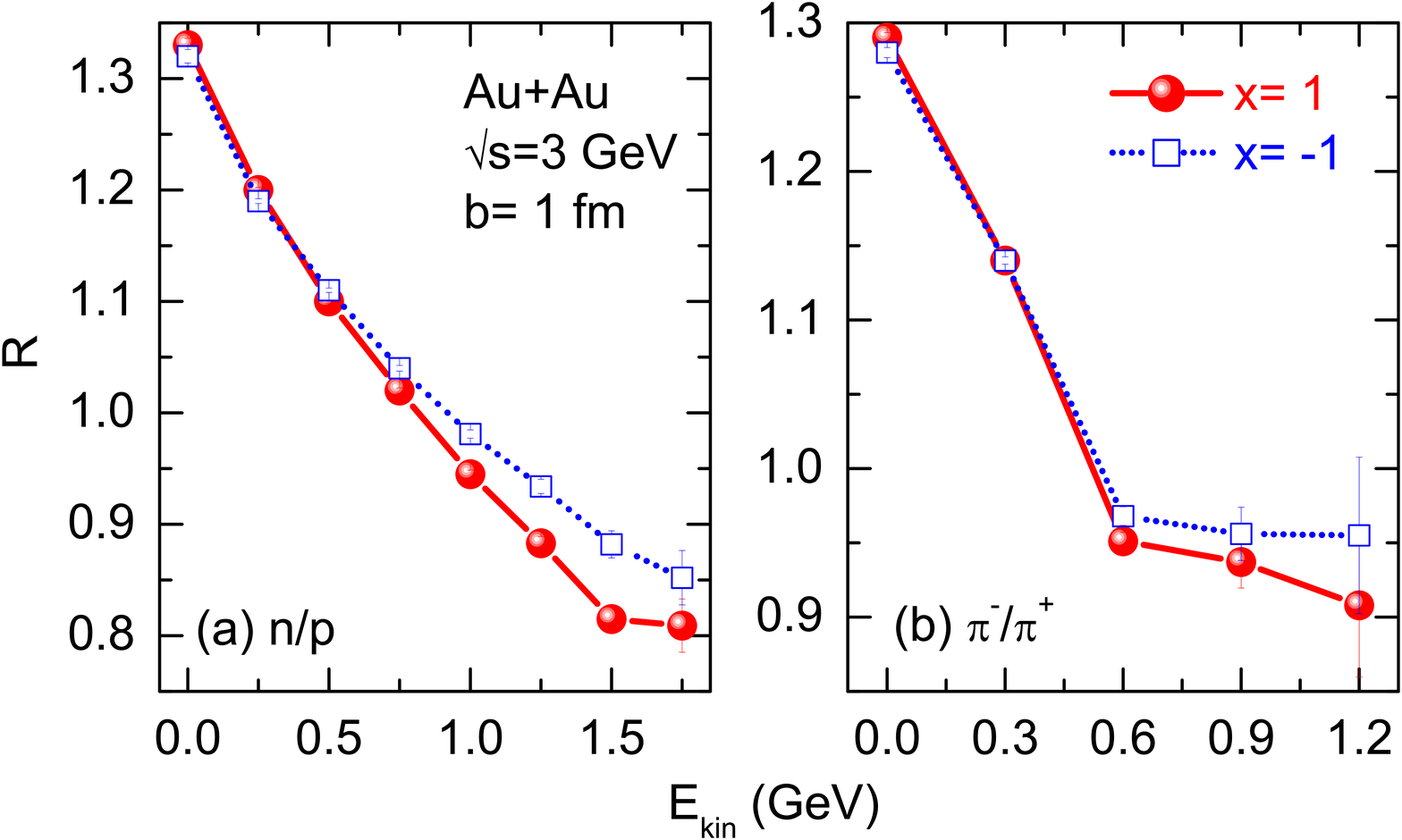}
  \includegraphics{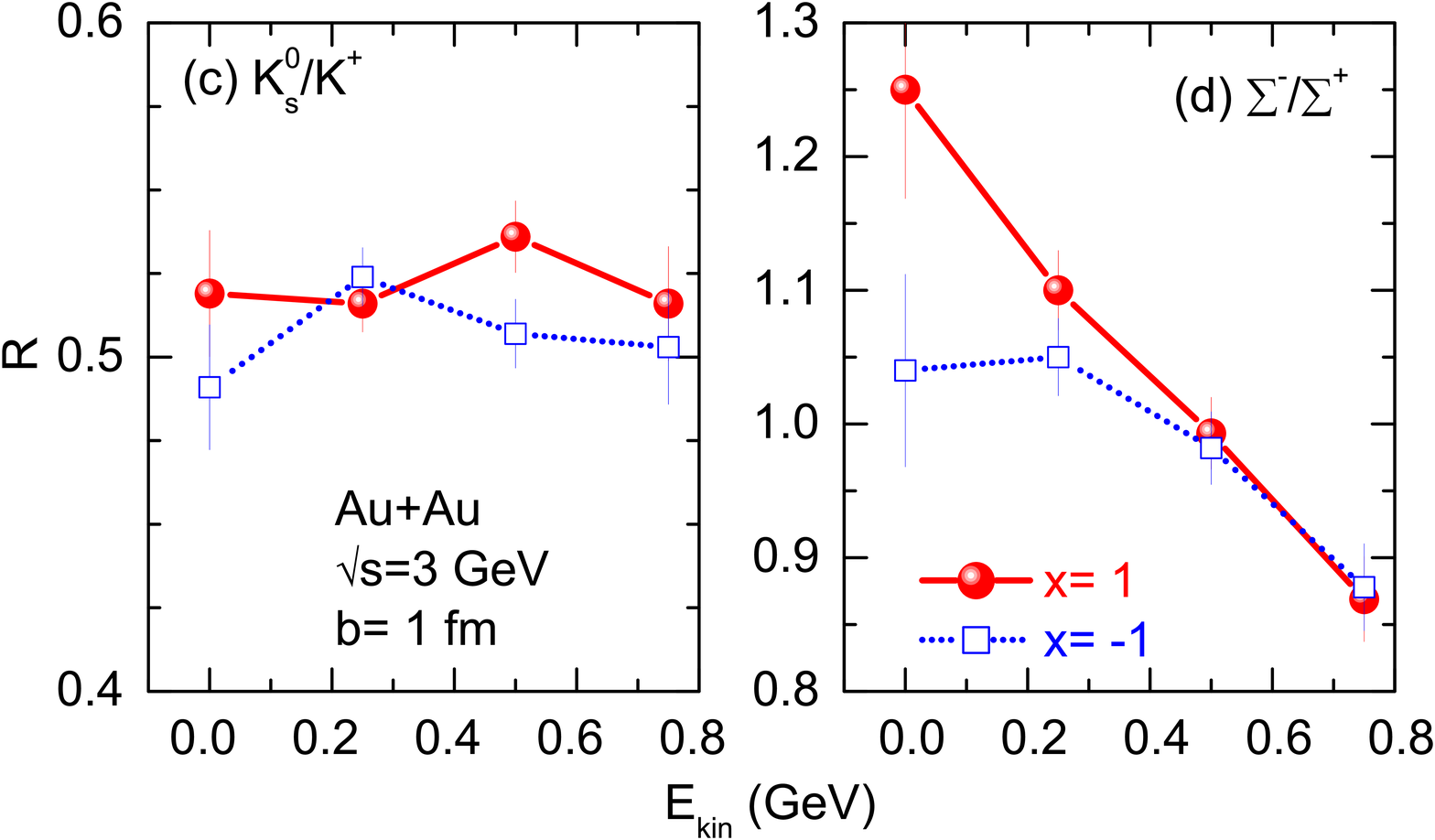}
    }
  \caption{The Kinetic energy distributions of $n/p$, $\pi^{-}/\pi^{+}$, $K_{s}^{0}/K^{+}$, and $\Sigma^{-}/\Sigma^{+}$ ratios in the central Au+Au reactions with stiff and soft symmetry energies at $\sqrt{s_{NN}}$ = 3 GeV.}\label{nonstrange}
\end{figure*}
\subsection{The $\Xi^{-}/\Xi^{0}$ ratio as a probe of \esym at suprasaturation densities}
The $\Xi^{-}$ and $\Xi^{0}$ yields and their ratio are shown as functions of time in panels (b) and (c) of Fig.~\ref{evo}. Both yields saturate at about t = 20 fm/c while their ratio already saturated much earlier around 10 fm/c. It is seen that the $\Xi$ production has evident isospin and symmetry energy effects. The yield of $\Xi^{-}$ with the soft symmetry energy is significantly larger than that with the stiff one. While for the yield of $\Xi^{0}$, the soft symmetry energy gives a somewhat smaller value compared with the stiff one. These isospin dependences can be understood from the most relevant chain reactions leading to the $\Xi^{-,0}$ production through the $\Lambda\Sigma^{-,+}\rightarrow\Xi^{-,0}$ reaction channels. Since pions involved in the $\pi^{-,+}N(n,p)\rightarrow\Sigma^{-,+}$ reactions are mostly from $nn(pp)\rightarrow\pi^{-,+}$ scatterings, one has effectively the dominating path $nn(pp)\rightarrow\Xi^{-,0}$. As discussed earlier and shown in the upper window of Fig.~\ref{evo}, a soft \esym at suprasaturation densities make the compressed region more neutron-rich, leading to more (less) $\Xi^- (\Xi^0)$ production there. Based on statistical models, one then expects a higher $\Xi^{-}/\Xi^{0}$ ratio. Indeed, as shown in Fig.~\ref{evo} (c), the soft \esym with x=1 leads to an about 30\% higher $\Xi^{-}/\Xi^{0}$ ratio compared to the stiff one (x=-1). This effect can also be seen clearly in both the kinetic energy and the transverse momentum spectra of the $\Xi^{-}/\Xi^{0}$ ratio as shown in Fig. \ref{rcas}. 

Although the $\Xi^{-}$ yield itself also shows significant sensitivity to the \esym as shown in the middle panel of Fig.~\ref{evo}, currently large uncertainties in calculating and measuring it may prevent us from using the $\Xi$ yield as a reliable EOS messenger. The significant advantages of using the $\Xi^{-}/\Xi^{0}$ ratio over the single $\Xi^{-}$ yield are the cancellations of some of the uncertainties (i.e., those related to the collision centrality and isoscalar parts of the $\Xi$ potentials). When using the $\Xi^{-}/\Xi^{0}$ ratio, besides the $\Xi^{-}$ that has been detected experimentally, one needs to reconstruct the $\Xi^{0}$ yield via the $\Xi^{0}$ $\rightarrow$ $\Lambda$ + $\pi^{0}$ ($\Lambda$ $\rightarrow$ p + $\pi^{-}$, $\pi^{0}$ $\rightarrow$ 2$\gamma$) channels.

\subsection{The $n/p$, $\pi^{-}/\pi^{+}$, $K_{s}^{0}/K^{+}$, and $\Sigma^{-}/\Sigma^{+}$ ratios}
As mentioned earlier, probing nuclear symmetry energy around $(1\sim 2)\rho_{0}$ by using heavy-ion collisions has been ongoing for the last two decades. The neutron to proton ratio $n/p$ of preequilibrium nucleons and the $\pi^{-}/\pi^{+}$ ratio are among the most studied observables \cite{BALI}. Therefore, it is interesting to investigate if these observables are still sensitive to the \esym in heavy-ion collisions at RHIC-BES energies and how they compare with the $\Xi^{-}/\Xi^{0}$ ratio discussed above. Fig.~\ref{nonstrange} shows the kinetic energy spectra of the $n/p$, $\pi^{-}/\pi^{+}$, $K_{s}^{0}/K^{+}$, and $\Sigma^{-}/\Sigma^{+}$ ratios with the same two \esym functions.
It is seen that while the $n/p$ and $\pi^{-}/\pi^{+}$ ratios at high kinetic energies still show appreciable \esym effects ($\sim 5\%)$, the effects are much smaller than those in heavy-ion reactions at lower beam energies \cite{BALI}.
From the isospin fractionation picture discussed earlier, with a stiff \esym at suprasaturation densities, the high (low) densities region is expected to be less (more) neutron rich, compared to the results with a soft \esym function. The final n/p ratio shown is for free nucleons. It is expected to be higher with the stiff \esym function. Interestingly, the $\pi^{-}/\pi^{+}$ ratio with the stiff symmetry energy is also higher than that with the soft one. Such behavior contradicts to that found at low beam energies. For example, in central Au+An reactions around $E_{\rm beam}/A=400\sim 1000$ MeV, a stiff \esym leads to a lower $\pi^{-}/\pi^{+}$ ratio reflecting directly the isospin asymmetry of the compressed phase \cite{Li2002,liyz2005}. At lower beam energies, most pions created are not energetic enough to induce secondary reactions to produce hyperons. Without passing the isospin information to hyperons, those pions survived the final state interactions do carry direct information about the isospin asymmetry of the compressed region. As we shall discuss next, this picture is changed when pions are main participants in creating strange mesons and hyperons at higher beam energies.

From the panel (c) and panel (d) in Fig.~\ref{nonstrange}, one sees that while the \esym effects on the $K_{s}^{0}/K^{+}$ ratio are not more than 6\% it affects the $\Sigma^{-}/\Sigma^{+}$ ratio as much as 20\% at low kinetic energies.
Since the $nn(pp)\rightarrow\Sigma^{-,+}$ dominates through the intermediate step $\pi+N (\rm {n~ or~ p})\rightarrow\Sigma$, the \esym information carried by the primordial pions is passed to the $\Sigma^{-}/\Sigma^{+}$ ratio. These $\Sigma$ hyperons mostly have low kinetic energies as a result of the reaction kinematics while the more energetic ones are produced through reactions involving two baryons that carry less isospin information than the primordial pions.
As we discussed earlier based on the baryon resonance or statistical model, the primordial $\pi^{-}/\pi^{+}$ ratio is
expected to be proportional to $(n/p)^2_{\rm like}$ of the participant region. The later is higher with the softer (x=1) \esym as shown in the middle-inset of Fig.~\ref{evo}. After the energetic pions have been converted to strange mesons and hyperons, the ones left over will have a reduced $\pi^{-}/\pi^{+}$ ratio as shown in the panel (b). While the opposite is expected for the newly produced particles through the scatterings involving pions. Indeed, the $K_{s}^{0}/K^{+}$, and $\Sigma^{-}/\Sigma^{+}$ ratios are higher with the softer (x=1) \esym as shown in the panels (c) and (d) of Fig.~\ref{nonstrange}.

The $K_{s}^{0}/K^{+}$ and $\Sigma^{-}/\Sigma^{+}$ ratios have been proposed as sensitive probes of the \esym since they are produced in the high density phase with little subsequent interactions \cite{ditoro,liq05}.
The relatively larger \esym effects on the $\Sigma^{-}/\Sigma^{+}$ ratio indicates that it might be a more suitable probe of the \esym at suprasaturation densities using heavy-ion reactions at RHIC-BES energies, provided the high energy neutral particles, neutrons and $\gamma$-rays, can be detected.

\section{Conclusions}
Using an updated ART model within the AMPT package by incorporating isospin and momentum-dependent hadronic mean fields, we explored observables that are sensitive to the high-density symmetry energy in $\sqrt{s_{NN}}$ = 3 GeV Au+Au reactions. We found that the doubly strange baryon ratio $\Xi^{-}/\Xi^{0}$ is particularly useful for the stated purpose
in comparison with the $n/p$, $\pi^{-}/\pi^{+}$, $K_{s}^{0}/K^{+}$, $\Sigma^{-}/\Sigma^{+}$ that has been proposed previously in heavy-ion collisions at lower beam energies.
Given the recent finding by the STAR Collaboration that the dense medium formed in such collisions is likely hadronic matter, the results of our study reported here are useful for probing the symmetry energy of dense hadronic matter near the onset of quark deconfinement.\\

%\section{Declaration of competing interest}
%
%The authors declare that they have no known competing financial interests or personal relationships that %could have appeared to influence the work reported in this paper.
%

\section{Acknowledgments}
This work is supported by the Strategic Priority Research Program of Chinese Academy of Sciences with Grant No. XDB34030000, the National Natural Science Foundation of China under Grant Nos. 11890712, 11875013, the Ministry of Science and Technology under Grant No. 2020YFE0202001 and the National Science Foundation under Grant No. PHY-2012947; The U.S. Department of Energy, Office of Science, under Award Number DE-SC0013702, the CUSTIPEN (China-U.S. Theory Institute for Physics with Exotic Nuclei) under the US Department of Energy Grant No. DE-SC0009971.

\end{document}